\documentclass[aps,pra,superscriptaddress,twocolumn]{revtex4}

\usepackage{amstext,amsmath,amssymb,ulem}
\usepackage{euscript}
\usepackage{color}

\usepackage[T1]{fontenc}
\usepackage[utf8]{inputenc}
\usepackage{graphicx}
\usepackage{mathptm}
\usepackage[colorlinks]{hyperref}

\newcommand{\beq}{\begin{equation}}
\newcommand{\eeq}{\end{equation}}
\newcommand{\ket} [1] {\vert #1 \rangle}
\newcommand{\bra} [1] {\langle #1 \vert}

\newcommand{\tr}{\mathop{\mathrm{tr}}}

\newcommand{\ba}{\begin{align}}
\newcommand{\ea}{\end{align}}
\newcommand{\bea}{\begin{eqnarray}}
\newcommand{\eea}{\end{eqnarray}}

\newcommand{\lmin}{l_{\tiny \mbox{min}}}
\newcommand{\lmax}{l_{\tiny \mbox{max}}}
\newcommand{\adis}{a'}
\newcommand{\awave}{a}
\newcommand{\Vdis}{{V'}}
\newcommand{\Vwave}{V}

\newcommand{\scale}{[\vec{s}]{\lmin}}
\newcommand{\scalezero}{[\vec{s}]0}
\newcommand{\wave}{[\vec{w}]l}
\newcommand{\scaler}[1]{[\vec{s}]#1}
\newcommand{\waver}[1]{[\vec{w}]#1}

\makeatletter
\@ifundefined{textcolor}{}
{%
 \definecolor{BLACK}{gray}{0}
 \definecolor{WHITE}{gray}{1}
 \definecolor{RED}{rgb}{1,0,0}
 \definecolor{GREEN}{rgb}{0,.4,0}
 \definecolor{BLUE}{rgb}{0,0,1}
 \definecolor{CYAN}{cmyk}{1,0,0,0}
 \definecolor{MAGENTA}{cmyk}{0,1,0,0}
 \definecolor{YELLOW}{cmyk}{0,0,1,0}
 }

\makeatother

\usepackage{bm}
\usepackage{mathtools}

\def\id{I}

\def\1{\mat{\id}}

\def\mat#1{\mathbf{#1}}

\renewcommand{\vec}[1]{\bm{\mathrm{#1}}}

\renewcommand{\sout}[1]{}

\begin{document} 
\title{Multi-scale quantum simulation of quantum field theory using wavelets}
\author{Gavin K. Brennen}
\affiliation{Centre for Engineered Quantum Systems, Department of Physics and Astronomy, Macquarie University, North Ryde, NSW 2109, Australia}
\author{Peter Rohde}
\affiliation{Centre for Engineered Quantum Systems, Department of Physics and Astronomy, Macquarie University, North Ryde, NSW 2109, Australia}
\author{Barry C. Sanders}
\affiliation{Institute for Quantum Science and Technology, University of Calgary, Alberta, Canada T2N 1N4}
\affiliation{Program in Quantum Information Science, Canadian Institute for Advanced Research, Toronto, Ontario M5G 1Z8, Canada}
\affiliation{Centre for Engineered Quantum Systems, Department of Physics and Astronomy, Macquarie University, North Ryde, NSW 2109, Australia}
\author{Sukhi Singh}
\affiliation{Centre for Engineered Quantum Systems, Department of Physics and Astronomy, Macquarie University, North Ryde, NSW 2109, Australia}

\begin{abstract}
A successful approach to understand field theories is to resolve the physics into different length or energy scales using the renormalization group framework.  
We propose a quantum simulation of quantum field theory which encodes field degrees of freedom in a wavelet basis---a multi-scale description of the theory. Since wavelets are compact wavefunctions, this encoding allows for quantum simulations to create particle excitations with compact support and provides a natural way to associate observables in the theory to finite resolution detectors. We show that the wavelet basis is well suited to compute subsystem entanglement entropy by dividing the field into contributions from short-range wavelet degrees of freedom and long-range scale degrees of freedom, of which the latter act as renormalized modes which capture the essential physics at a renormalization fixed point.

\end{abstract}


\maketitle

\section{Introduction}
\label{intro}
Wavelets are a versatile basis to represent functions which are neither localised in position or momentum. They are best known for their use in signal processing such as in the Joint Photographic Experts Group (JPEG) compression where they can represent and compress data at multiple spatial scales~\cite{Mallet} with low loss of fidelity. They are also being adopted to speed up calculations for a plethora of problems in science including quantum molecular dynamics~\cite{TW1997}, density functional theory~\cite{Natarajan}, and Monte Carlo simulations on lattice~\cite{Ismail}, which are of enormous importance for quantum chemistry, solid state, and statistical physics. 
Further there are potential applications to high energy physics where a wavelet basis been proposed as a way to regularize quantum field theories~\cite{Altaisky2007}.

At the same time that these advances have been made in classical computations, algorithms have been developed to attack difficult problems in quantum mechanics by using quantum simulators~\cite{Nori}. However, most quantum algorithms for simulation of dynamics in real space use some version of bases which are localised in position and/or momentum and mapped into each other by Fourier transforms. While the quantum Fourier transform is efficient, more efficient evolutions may be possible for quantum states which are not localized in either basis.

In Ref.~\cite{JLP2012}, the authors provide a quantum algorithm to simulate scalar bosonic field theories which achieves accurate estimation of scattering matrix probabilities in a time exponentially faster than known classical algorithms.  
Here we present a wavelet based quantum simulation. A key feature of this basis choice is that we need not discretize space, rather we choose a representative scale to capture features of the wave function and can add smaller scale features in a controlled manner.   There are several advantages to using wavelets in the context of quantum simulation algorithms for quantum field theory.  First,
wavelets have a built in scaling structure which could be used to compute expectation values of operators such as energy density and two point correlations functions at different length scales. This information could then be used to compute fixed points of renormalisation flows~\cite{waveletsandrenorm}.  Second,
the wavelet basis has a well defined procedure to include local gauge invariance via covariant derivatives at every length scale~\cite{AK2013,BP2013}.  Third, in the spirit of quantum information, a wavelet basis is a natural one to to describe quantum fields by the scale of a measurement.  This can obviate issues with divergences of Greens functions that arise in calculations using point like operators~\cite{Altaisky2010}.

The wavelet basis consists of ``scale functions'' at a given length scale and ``wavelet functions'' at finer length scales. The scale functions are scale-invariant by construction and thus it is not surprising that they turn out to span the subspace that captures the essential physics at the renormalization fixed point. Resolving the description of a system according to length scale has also led to a successful numerical approach---the Multi-scale Renormalization Ansatz (MERA)~\cite{dMERA}---primarily for classical simulation of both discrete quantum many-body systems and also field theories~\cite{cMERA} (in the latter case, the success of the ansatz has been demonstrated for free field theories).

We first briefly introduce in Sec. \ref{Sec1} the essential features of wavelets focusing on a particular family, the Daubechies wavelets, which are related to each other by dyadic scaling and discrete translations. In Sec. \ref{Sec2} we represent the Hamiltonian for a scalar bosonic field theory in $d=1$ spatial dimension in a wavelet basis with straightforward extension to higher $d$. We show how to encode the ground state of the free field theory in a register of qubits or bosonic modes and how to create single particle excitations and turn on quartic interactions. The complexity of this simulation is similar to the algorithm of~\cite{JLP2012} that uses the discretized position basis as discussed in Sec. \ref{Sec4}. In Sec. \ref{Sec3} we demonstrate how encoding the free field ground state in the wavelet basis captures the essential physics of field theories from an entanglement perspective. The logarithmic scaling of ground state entanglement in the massless case is entirely captured by the coarse scale degrees of freedom, indicating that these are indeed a representation of the renormalized degrees of freedom of theory. Our results are summarized in the conclusions.

 \section{Basic properties of Daubechies wavelets}
 \label{Sec1}
Wavelets constitute an orthonomal basis for the Hilbert space $L^2(\mathbb{R})$ of square integrable functions on the line and we briefly review some of their properties here. For a comprehensive survey see Ref.~\cite{Mallet}.  Generically, wavelets  are defined in terms of a mother wavelet function $w(x)$ and a father scaling function $s(x)$ by taking linear combinations of shifts and rescalings thereof.   For the remainder we focus on one family known as Daubechies $\mathcal{K}$-wavelets where the role of $\mathcal{K}\in\mathbb{Z}^+$ will be described below.  First we introduce two unitary operators on $L^2(\mathbb{R})$:  $\mathcal{T}$ for  discrete translation and $\mathcal{D}$ for scaling defined by the action on a function $f\in L^2(\mathbb{R})$:
\begin{equation}
\mathcal{D}f(x)=\sqrt{2}f(2x);\quad \mathcal{T}f(x)=f(x-1).
\end{equation}
The father scaling function $s(x)$ is a solution to the linear renormalisation group equation
\begin{equation}\label{Father}
s(x)=\mathcal{D}\left[\sum_{n=0}^{2\mathcal{K}-1}h_n\mathcal{T}^n s(x)\right],
\end{equation}
reading, \emph{first block average then rescale}.  The $2\mathcal{K}$ real coefficients $\{h_n\}$ are computed analytically for $\mathcal{K}< 4$ and are solved for numerically otherwise. Given the solution to $s(x)$, scale $2^{-k}$ scaling functions are defined by applying $n$ unit translations followed by $k$ scaling transformations on the father:
\begin{equation}
s^k_n(x) =D^k T^n s(x).
\end{equation}

\begin{figure}[t]
\begin{center}
\includegraphics[width=8cm]{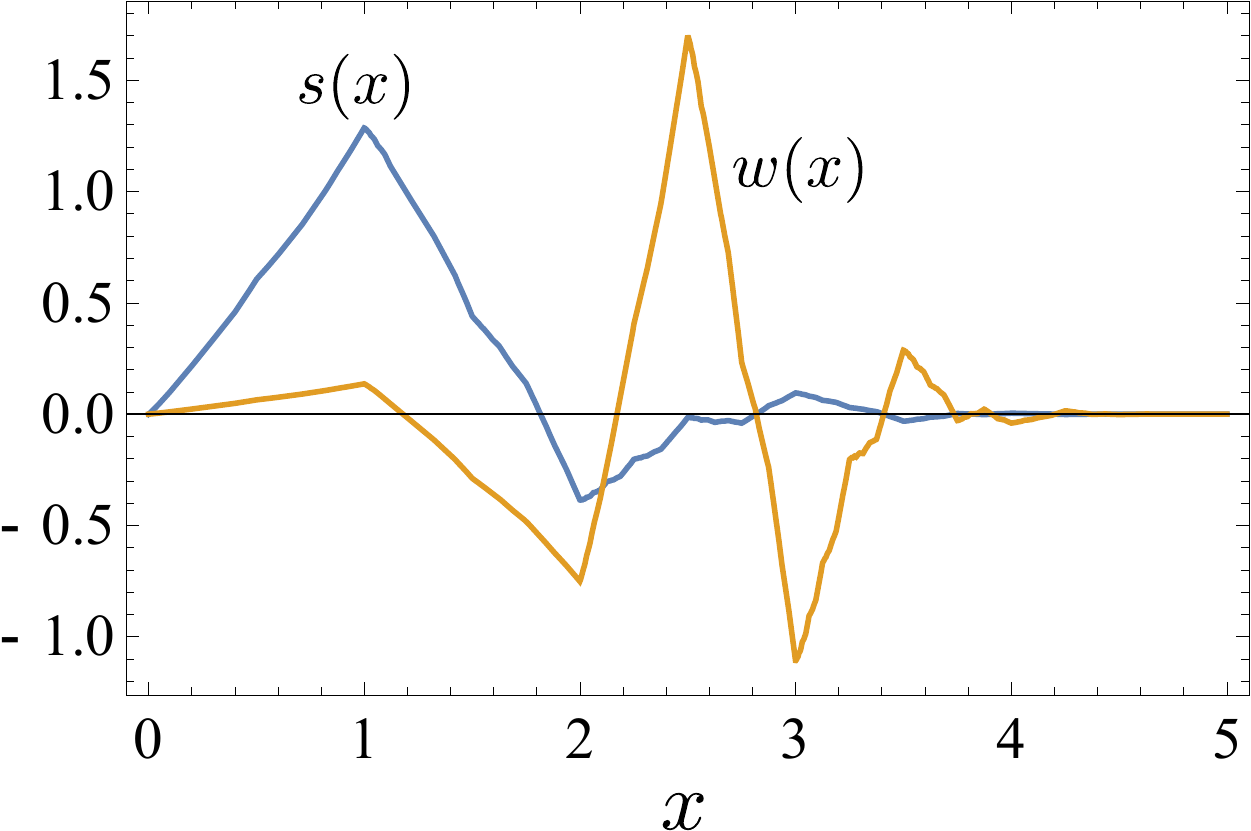}
\end{center}
\caption{The father scaling function $s(x)$ and mother wavelet $w(x)$ with support on $[0,5]$ for the Daubechies $\mathcal{K}=3$ wavelet family.  The functions have continuous first derivatives.}
\label{fig:scatterers}
\end{figure}

The scaling functions are normalised so that
\begin{equation}
\int dx\ s^k_n(x)=1.
\end{equation}
The mother wavelet $w(x)$ and the father $s(x)$ have the property that they are neither localised in position or momentum.   The wavelets take the following form:
\begin{equation}
w(x)=\sum_{n=0}^{2\mathcal{K}-1}g_n\mathcal{D}\mathcal{T}^n s(x)=\sum_{n=0}^{2\mathcal{K}-1}g_ns^1_n(x),
\end{equation}
where the set of coefficients $\{g_n\}$ are obtained from $\{h_n\}$ by reversing the order and alternating signs:  $g_n=(-1)^nh_{2\mathcal{K}-1-n}$.  Scale $2^{-k}$ wavelets are obtained by translating and scaling the mother:
\begin{equation}
w^k_n(x)=\mathcal{D}^k \mathcal{T}^n w(x).
\end{equation}
The index $\mathcal{K}$ specifies the number of vanishing moments of the wavelets, i.e. 
\[
\int\ dx\ w(x)x^p=0\quad p=0,..,\mathcal{K}.
\]
The vanishing of the zeroth moment is synonymous with the admissibility condition which guarantees that the wavelet basis is square integrable~\cite{Mallet}.
Choosing larger $\mathcal{K}$ means more features can be captured at a given scale, however at the expense of additional computational cost since more translations are needed during block averaging. Daubechies wavelets are optimal in the sense that they have the smallest size support for a given number of vanishing moments~\cite{Mallet}. 
The basis functions $s^k_n(x)$ and $w^k_n(x)$ have support on $[2^{-k}n,2^{-k}(n+2\mathcal{K}-1)]$ and satisfy the following orthonormality relations:
\begin{equation}
\begin{split}
&\int dx\ s^k_n(x)s^k_m(x)=\delta_{m,n},\\
&\int dx\ s^k_n(x)w^{k+l}_m(x)=0\quad (l\geq 0),\\
&\int dx\ w^k_n(x)w^l_m(x)=\delta_{m,n}\delta_{k,l}.
\end{split}
\end{equation}
By the last relation, the wavelets constitute normalised wave functions.  The scaling functions at scale $2^{-k}$ are complete in that
\begin{equation}
\sum_{n=-\infty}^{\infty} \frac{1}{\sqrt{2^k}}s^k_n(x)=1.
\end{equation}
A final important property of the Daubechies $\mathcal{K}$-wavelets is that they are $\mathcal{K}-2$ times differentiable. 

Linear superpositions of functions $\{s^k_n(x)\}_{n=-\infty}^{\infty}$ (with square summable coefficients) span a subspace $\mathcal{H}_k$ of $L^2(\mathbb{R})$ which is the scale $2^{-k}$ subspace and which is a proper subspace of a smaller scale space $\mathcal{H}_k\subset \mathcal{H}_{k+m}\ (m>0)$.  Linear combinations of the scale $2^{-k}$ wavelet functions $\{w^k_n(x)\}_{n=-\infty}^{\infty}$ span the orthocomplement $\mathcal{W}_k$ of $\mathcal{H}_k$ in $\mathcal{H}_{k+1}$:
$\mathcal{H}_{k+1}=\mathcal{H}_{k}\oplus\mathcal{W}_{k}$.  We can use a set of scaling functions $\{s^k_n(x)\}_{n=-\infty}^{\infty}$ to represent features down to scale $2^{-k}$ and a set of wavelets $\{w^k_n(x)\}_{n=-\infty}^{\infty}$ to represent features down to scale $2^{-(k+1)}$ that cannot be represented at scale $2^{-k}$.
The whole space has the following decomposition satisfied for \emph{any} finite $k$:
\begin{equation}
L^2(\mathbb{R})=\mathcal{H}_{k}\bigoplus_{l=k}^{\infty}\mathcal{W}_{l},
\label{Hilbertspace}
\end{equation}
meaning that for a fixed scale $2^{-k}$ the set 
\[
\{s^k_n(x)\}_{n=-\infty}^{\infty}\bigcup \{w^l_n(x)\}_{n=-\infty,l=k}^{\infty,\infty}
\]
 span a basis for $L^2(\mathbb{R})$. 

\section{A wavelet representation of quantum fields}
\label{Sec2}
\subsection{Free field ground state represented in the wavelet basis}
The class of theories we address are the scalar (massive or massless) bosonic $\hat{\Phi}^4$ theory in $d\in{\mathbb N}$ spatial dimensions. These are given by the Hamiltonian:
\begin{equation}
\hat{H}=\hat{H}^{(0)}+\hat{H}^{(I)},
\end{equation}
where the free field contribution is
\begin{equation}
\hat{H}^{(0)}=\int d^dx\ \frac{1}{2}(\hat{\Pi}^2(\bm{x},t)+\left(\vec{\nabla}\hat{\Phi}^2(\bm{x},t))+m_0^2\hat{\Phi}^2(\bm{x},t)\right),
\label{Ham}
\end{equation}
and the interaction term is
\begin{equation}
\hat{H}^{(I)}=\int d^dx\ \frac{\lambda_0}{4!}\hat{\Phi}^4(\bm{x},t).
\end{equation}
The canonical momentum is 
\begin{equation}
\hat{\Pi}(\bm{x},t)=\frac{\partial \hat{\Phi}(\bm{x},t)}{\partial t},
\end{equation}
which together with the field are normalised to satisfy the equal time commutation relation $[\hat{\Phi}(\bm{x},t),\hat{\Pi}(\bm{y},t)]=i\delta^d(\bm{x}-\bm{y})$ $(\hbar\equiv 1)$.
Here the phase velocity of waves in this theory is set so that the speed of light is $1$, the bare mass is $m_0$, and the strength of the interactions is dictated by $\lambda_0$.

To apply wavelets to the field theory we follow the prescription given in Ref.~\cite{BP2013}. Because the Hamiltonian involves terms with no higher than first derivatives, it suffices to choose the Daubechies $\mathcal{K}=3$ wavelet family which have continuous first derivatives for the scale and wavelet functions.  This will guarantee that we have analytic forms for the coupling matrix elements in the wavelet basis while also providing for a minimal size support for the functions, a feature which reduces the number of non-zero coupling terms that appear in the Hamiltonian.  We present the $d=1$ case as it makes the notation considerably simpler and captures the salient features of the algorithm. The wavelet representation can easily be extended to higher dimensions (see Appendix \ref{higherD}) using a cartesian product of wavelets and scale functions. First we decompose the field and its conjugate in the wavelet basis as:
\begin{equation}
\begin{split}
\hat{\Phi}(x,t)&=\sum_{n\in \mathbb{Z}} \hat{\Phi}^{\scale}(n,t)s^{\lmin}_n(x)+\sum_{n\in \mathbb{Z}}\sum_{l=\lmin}^{\infty}\hat{\Phi}^{\wave}(n,t)w^l_{n}(x),\\
\hat{\Pi}(x,t)&=\sum_{n\in \mathbb{Z}} \hat{\Pi}^{\scale}(n,t)s^{\lmin}_n(x)+\sum_{n\in \mathbb{Z}}\sum_{l=\lmin}^{\infty}\hat{\Pi}^{\wave}(n,t)w^l_{n}(x),
\end{split}
\end{equation}
where the coarsest scale in the theory corresponds to $2^{-\lmin}$. Henceforth, we drop the dependence of the fields and their conjugates on time.  The discrete field operators are projections of the field operators onto the scaling and wavelet functions (here $l\geq \lmin$):
\begin{equation}
\begin{split}
\hat{\Phi}^{\scale}(n)&=\int dx\ \hat{\Phi}(x)s^{\lmin}_{n}(x),~~
\hat{\Phi}^{\wave}(n)=\int dx\ \hat{\Phi}(x,t)w^l_{n}(x),\\
\hat{\Pi}^{\scale}(n)&=\int dx\ \hat{\Pi}(x)s^{\lmin}_{n}(x),~~
\hat{\Pi}^{\wave}(n)=\int dx\ \hat{\Pi}(x)w^l_{n}(x),
\end{split}
\end{equation}
and they satisfy the following equal time commutation relations (assuming here that $\lmin\leq r,s$):
\begin{equation}
\begin{split}
\ [\hat{\Phi}^{\scale}(n),\hat{\Phi}^{\scale}(m)]&=0,\quad [\hat{\Pi}^{\scale}(n),\hat{\Pi}^{\scale}(m)]=0,\\
\ [\hat{\Phi}^{\scale}(n),\hat{\Pi}^{\scale}(m)]&= i \delta_{n,m},\\
\ [\hat{\Phi}^{\waver{r}}(n),\hat{\Phi}^{\waver{s}}(m)]&=0,\quad
[\hat{\Pi}^{\waver{r}}(n),\hat{\Pi}^{\waver{s}}(m)]=0,\\
\ [\hat{\Phi}^{\waver{r}}(n),\hat{\Pi}^{\waver{s}}(m)]&=i\delta_{r,s}\delta_{n,m},\\
\ [\hat{\Phi}^{\waver{r}}(n),\hat{\Phi}^{\waver{s}}(m)]&=0,\quad [\hat{\Pi}^{\waver{r}}(n),\hat{\Pi}^{\waver{s}}(m)]=0\\
\ [\hat{\Phi}^{\waver{r}}(n),\hat{\Pi}^{\waver{s}}(m)]&=0,\quad [\hat{\Pi}^{\waver{r}}(n),\hat{\Phi}^{\waver{s}}(m)]=0.
 \end{split}
 \end{equation}
 The discrete annihilation operators, for the scaling and wavelet fields respectively, are
\begin{equation}
 \begin{split}
 \hat{a}^{\lmin}(n)&=\frac{1}{\sqrt{2}}\left(\sqrt{\gamma(\lmin)} \hat{\Phi}^{\scale}(n)+i\frac{1}{\sqrt{\gamma(\lmin)}}\hat{\Pi}^{\scale}(n)\right),\\
  \hat{b}^r(n)&=\frac{1}{\sqrt{2}}\left(\sqrt{\gamma(r)} \hat{\Phi}^{\waver{r}}(n)+\frac{i}{\sqrt{\gamma(r)}}\hat{\Pi}^{\waver{r}}(n)\right),
  \end{split}
\end{equation}
 and the inverse relations are
 \begin{equation}
 \begin{split}
\hat{\Phi}^{\scale}(n) &= \frac{1}{\sqrt{2\gamma^{[\vec{s}]}(\lmin)}} \left( \hat{a}^{{\lmin}\dagger}(n)+\hat{a}^{\lmin}(n)\right),\\
\hat{\Pi}^{\scale}(n) &= i\sqrt{\frac{\gamma^{[\vec{s}]}(\lmin)}{2}}\left(\hat{a}^{{\lmin}\dagger}(n)-\hat{a}^{\lmin}(n)\right),\\
\hat{\Phi}^{\waver{r}}(n) &= \frac{1}{\sqrt{2\gamma^{[\vec{w}]}(r)}} \left(\hat{b}^{r\dagger}(n)+\hat{b}^{r}(n)\right),\\
\hat{\Pi}^{\waver{r}}(n) &= i\sqrt{\frac{\gamma^{[\vec{w}]}(r)}{2}} \left(\hat{b}^{r\dagger}(n)-\hat{b}^{r}(n)\right).
  \end{split}
  \end{equation}  
  
Each annihilates the free field vacuum and together with the set of adjoint creation operators they satisfy the bosonic commutation relations:
  \begin{equation}
  \begin{split}
  \ [\hat{a}^{\lmin}(n),\hat{a}^{{\lmin}\dagger}(m)]&=\delta_{m,n},\\
  \ [\hat{b}^l(n),\hat{b}^{j \dagger}(m)]&=\delta_{m,n}\delta_{j,l},
  \end{split}
  \end{equation}
 with all others commutators vanishing.  The Hilbert space for the free field theory is spanned by linear combinations of products of the creation operators from the set $a^{{\lmin}\dagger}(m),b^{l\dagger}(m)$ applied to $\ket{G}$.
 
 The coefficients $\gamma$ depend on the scale $2^{-{\lmin}}$ and the mass $m_0$ as follows:
 \begin{equation}
 \begin{split}
\gamma^{[\vec{s}]}({\lmin})&=\frac{1\pm\sqrt{1-4\nu^{[\vec{s}]}({\lmin})\eta^{[\vec{s}]}({\lmin})}}{2\nu^{[\vec{s}]}({\lmin})},\\
\gamma^{[\vec{w}]}(r)&=\frac{1\pm\sqrt{1-4\nu^{[\vec{w}]}(r)\eta^{[\vec{w}]}(r)}}{2\nu^{[\vec{w}]}(r)},
 \label{gammas}
 \end{split}
 \end{equation} 
where $\ket{G}$ is the free field vacuum state,
\begin{equation}
\begin{split}
\nu^{[\vec{s}]}({\lmin}) &= \bra{G}\hat{\Phi}^{\scaler{\lmin}}(0)\hat{\Phi}^{\scaler{\lmin}}(0)\ket{G},\\
\nu^{[\vec{w}]}(r) &= \bra{G}\hat{\Phi}^{\waver{r}}(0)\hat{\Phi}^{\waver{r}}(0)\ket{G},\\
\eta^{[\vec{s}]}(\lmin) &= \bra{G}\hat{\Pi}^{\scaler{\lmin}}(0)\hat{\Pi}^{\scaler{\lmin}}(0)\ket{G},\\
\eta^{[\vec{w}]}(r) &= \bra{G}\hat{\Pi}^{\waver{r}}(0)\hat{\Pi}^{\waver{r}}(0)\ket{G},
\end{split}
\end{equation}and the $\pm$ sign is chosen according to the case that makes the expression positive. 
In order to obtain these factors we need to explicitly compute the expectation value of quadratic products of the discrete field operators in the ground state which are as follows:
\begin{equation}
\begin{split}
\nu^{[\vec{s}]}({\lmin})&=
\frac{1}{(2\pi)^3}\int\ dxdydp\frac{s^{\lmin}_0(x)s^{\lmin}_0(y)}{2\omega(\vec{p})}e^{i\vec{p}\cdot (x-y)},\\
\eta^{[\vec{s}]}({\lmin})&=
\frac{1}{(2\pi)^3}\int\ dxdydp\frac{s^{\lmin}_0(x)s^{\lmin}_0(y)\omega(\vec{p})}{2}e^{i\vec{p}\cdot (x-y)},\\
\nu^{[\vec{w}]}(r)&=
\frac{1}{(2\pi)^3}\int\ dx dy dp\frac{w^r_{\vec{0}}(x)w^r_{\vec{0}}(y)}{2\omega(\vec{p})}e^{i\vec{p}\cdot (x-y)},\\
\eta^{[\vec{w}]}(r)&=
\frac{1}{(2\pi)^3}\int\ dx dy dp\frac{w^r_{\vec{0}}(x)w^r_{\vec{0}}(y)\omega(\vec{p})}{2}e^{i\vec{p}\cdot (x-y)},
\end{split}
\label{overlaps}
\end{equation}
where $\omega(\vec{p})=\sqrt{m_0^2+\vec{p}^2}$,
is the single particle energy.  All the terms in Eq. \ref{overlaps} can be computed numerically for a given input mass $m_0$. 

Following Ref.~\cite{BP2013} we decompose the free field Hamiltonian into three pieces 
\begin{equation}\label{FreeHam}
\hat{H}^{(0)}=\hat{H}_{\rm ss}+\hat{H}_{\rm ww}+\hat{H}_{\rm sw}.
\end{equation}
We fix a scale $2^{-\lmin}$ so that the Hilbert space is decomposed as in Eq. \ref{Hilbertspace}. Then the constituent terms of the Hamiltonian are
\begin{equation}\label{Haminwaveletbasis}
\begin{split}
\hat{H}_{\rm ss}&=\frac{1}{2}\big(\sum_{n\in\mathbb{Z}} :\hat{\Pi}^{\scale}(n)\hat{\Pi}^{\scale}(n):\\
&+m_0^2\sum_{n\in \mathbb{Z}} :\hat{\Phi}^{\scale}(n)\hat{\Phi}^{\scale}(n):\\
&+\sum_{m,n\in\mathbb{Z}}:\hat{\Phi}^{\scale}(m)D^k_{m,n}\hat{\Phi}^{\scale}(n):\big),\\
\hat{H}_{\rm ww}&=\frac{1}{2}\big(\sum_{n\in\mathbb{Z}}\sum_{l\geq \lmin} :\hat{\Pi}^{\wave}(n)\hat{\Pi}^{\wave}(n):\\ &+m_0^2\sum_{n\in\mathbb{Z}}\sum_{l\geq \lmin} :\hat{\Phi}^{\wave}(n)\hat{\Phi}^{\wave}(n):\\
&+\sum_{m,n\in\mathbb{Z}}\sum_{l,j\geq \lmin}:\hat{\Phi}^{\wave}(m)D^{l,j}_{m,n}\hat{\Phi}^{\waver{j}}(n):\big),\\
\\
\hat{H}_{\rm sw}&=\frac{1}{2}\sum_{m,n\in \mathbb{Z}}\sum_{l\geq \lmin}:\hat{\Phi}^{\wave}(m)D^{l,{\lmin}}_{m,n}\hat{\Phi}^{\scale}(n):,
\end{split}
\end{equation}
where $:\hat{O}:$ indicates normal ordering of the operator $\hat{O}$ is taken.  The operator $\hat{H}_{\rm ss}$ describes physics at a scale $2^{-{\lmin}}$ involving interactions between scale field degrees of freedom, $\hat{H}_{\rm ww}$ describes physics at a finer scales $2^{-l}$ for $(l>{\lmin})$ involving interactions between wavelet degrees of freedom, and $\hat{H}_{\rm sw}$ describes coupling between scale fields at resolution $2^{-{\lmin}}$ and finer wavelet degrees of freedom.  While there are an infinite number of finer scale degrees of freedom we truncate to ${\lmax}$ consistent with momentum cutoffs in physical theories.  Specifically, the maximum momentum for a single particle excitation is $p_{\rm max}\simeq 2^{{\lmax}}$ as described in Sec. \ref{momentum}. The coupling coefficients are 
\begin{equation}
\begin{split}
D^{\lmin}_{m,n}&=\int dx\ \nabla s^{\lmin}_{m}(x)\cdot \nabla s^{\lmin}_{n}(x),\\
D^{l,j}_{m,n}&=\int dx\ \nabla w^l_{m}(x)\cdot \nabla w^j_{n}(x),\\
D^{l,{\lmin}}_{m,n}&=2 \int dx\ \nabla w^l_{m}(x)\cdot \nabla s^{\lmin}_{n}(x).\\
\end{split}
\end{equation} 
Many of these coefficients are computed in~\cite{BP2013} for the Daubuchies $\mathcal{K}=3$ wavelets. The choice of $\mathcal{K}=3$ ensures a continuous first derivative of the scaling functions which allows for computing these overlaps exactly.  Because the functions have compact support, the coefficients vanish unless $|n-m|\leq 4$.

Let the physical one dimensional volume be $L \awave$ where $L\in \mathbb{N}$ and $\awave$ is the unit of length at the base scale. The size of $L$ will be determined by the long wavelength physics that one wishes to capture.  At smaller scales, $2^{-l}$, the unit of length is $\awave2^{-l}$. We will work in normalised length units such that $\awave=1$, and we choose our base scale so that ${\lmin}=0$ such that the support of the scaling function $s^0_0(x)=s(x)$ is the interval $[0,5]$.  A plot of these functions is shown in Fig. \ref{fig:2}.
Now let us introduce notation for basis vectors in the wavelet basis. Basis vectors $\ket{r}_{w^j_{m}}$ denote amplitude $r$ in the wavelet mode ${w^j_{m}}$ such that $\hat{\Phi}^{\waver{j}}(m)\ket{r}_{w^j_{m}}=r\ket{r}_{w^j_{m}}$, and similarly $\hat{\Phi}^{\scalezero}(m)\ket{r}_{s^0_{m}}=r\ket{r}_{s^0_{m}}$.  We adopt a simplified notation for states in the tensor product space of the 
\begin{equation}
	V=L2^{{\lmax}+1}
\label{nummodes}
\end{equation}
modes utilizing the vector $\bm{r}=(r_0,\ldots r_{V-1})^T$ with
\begin{equation}\label{modeops}
\begin{array}{lll}
\ket{\vec{r}}&=&\ket{r_0}_{s^0_0}\otimes\cdots\otimes \ket{r_{L-1}}_{s^0_{L-1}}\otimes \ket{r_L}_{w^0_0}\otimes\cdots\otimes \ket{r_{2L-1}}_{w^0_{L-1}}\otimes\\
&&\ket{r_{2L}}_{w^1_0}\otimes\cdots\otimes \ket{r_{4L-1}}_{w^1_{2L-1}}\otimes \ket{r_{4L}}_{w^2_{0}} \cdots\otimes \ket{r_{V-1}}_{w^{{\lmax}}_{L2^{{\lmax}}-1}}.
\end{array}
\end{equation}


\begin{figure}[t]
\begin{center}
\includegraphics[width=\columnwidth]{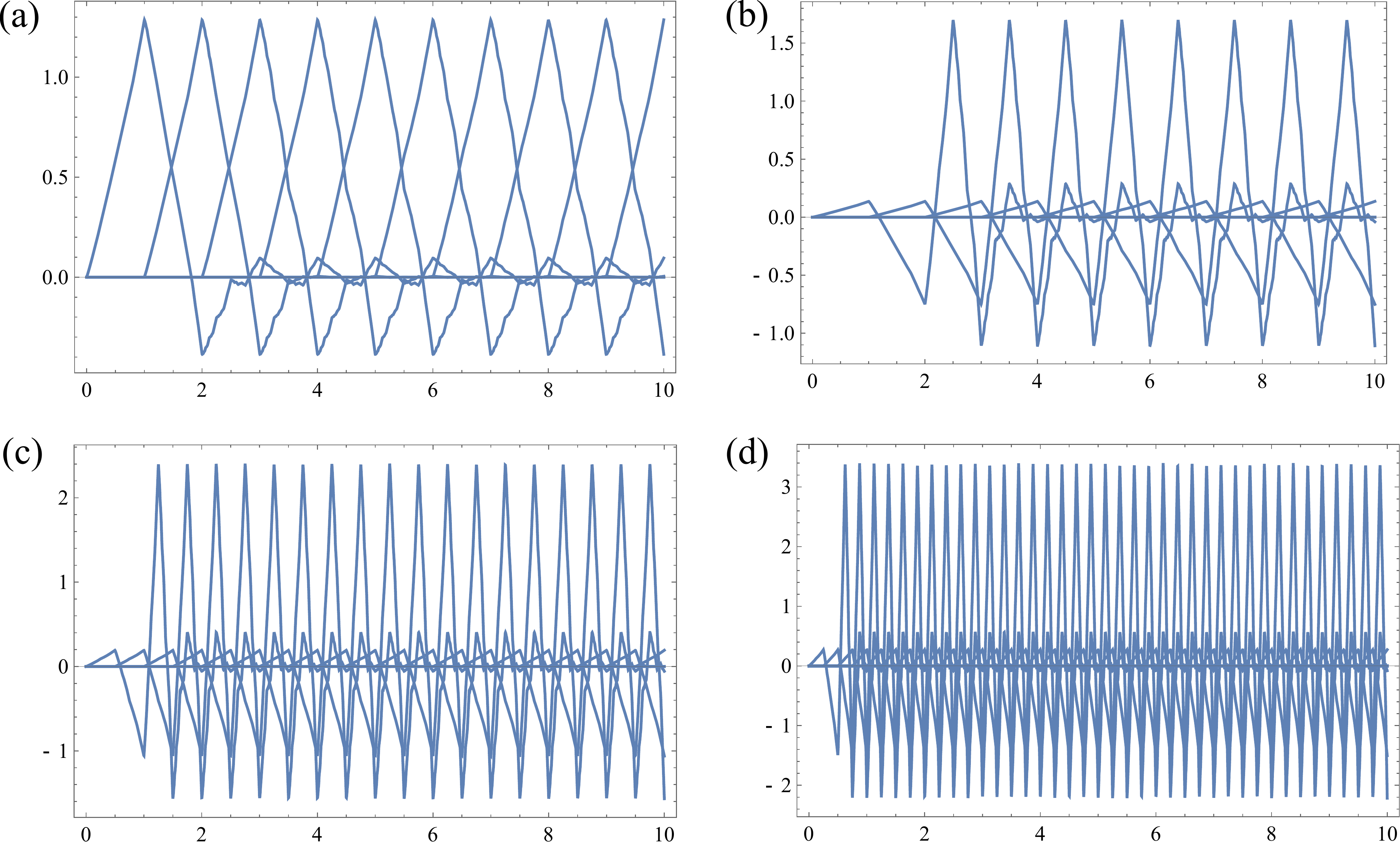}
\label{fig:2}
\end{center}
\caption{One dimensional Daubechies $\mathcal{K}=3$ scale functions and wavelets plotted as a function of $x$  at three scales  for a system of size $L=10$.  (a) Scale functions $\{s^0_{n}(x)\}_{n=0}^{L-1}$; (b) Wavelets $\{w^0_{n}(x)\}_{n=0}^{L-1}$; (c) $\{w^1_{n}(x)\}_{n=0}^{2L-1}$; and (d) $\{w^2_{n}(x)\}_{n=0}^{4L-1}$. Here and in the main text we assume hard wall boundaries.}
\label{fig:scatterers}
\end{figure}

The ground state of the free field theory $\hat{H}^{(0)}$ is then approximated by
\begin{equation}\label{waveletvac}
\ket{G}\approx \mathcal{N}^{-1}\int dr_0\ldots \int dr_{V-1}\ e^{-\frac{1}{2}\bm{r}^T K^{1/2}\bm{r}}\ket{\vec{r}},
\end{equation}
where the normalisation is $\mathcal{N}^{-1}=\det(K^{1/2})^{1/4}/\pi^{V/4}$.  Here the coupling matrix is
\begin{equation}\label{Kmatrix}
K=\left[\begin{array}{llll}[K_{ss}] & [K_{sw}(0)] & \cdots & [K_{sw}({\lmax})] \\
\ [K_{sw}(0)]^T  & [K_{ww}(0,0)]  & \ldots & [K_{ww}(0,{\lmax})]  \\
\vdots &  & \ddots &  \\
\ [K_{sw}({\lmax})]^T  & [K_{ww}(0,{\lmax})]^T & \cdots & [K_{ww}({\lmax},{\lmax})]
\end{array}\right].
\end{equation}
The scale-scale mode couplings are encoded in $K_{ss}$, the scale-wavelet couplings in $K_{sw}$ and the wavelet-wavelet couplings in $K_{ww}$. These matrices are:
\begin{equation}\label{BigK}
\begin{split}
&[K_{ss}]_{a,b}=m_0^2\delta_{a,b}+D^0_{a,b}
\quad \quad (0\leq a,b<L)\\
&[K_{sw}(l)]_{a,b}=D^{0,l}_{a,b}
\quad \quad (0\leq a<L, 0\leq b < L2^l, 0\leq l\leq {\lmax})\\
&[K_{ww}(l,j)]_{a,b}=m_0^2\delta_{a,b}\delta_{j,l}+D^{l,j}_{a,b}\\
&\quad \quad (0\leq a<L2^l, 0\leq b < L2^j, 0\leq j\leq l\leq {\lmax})\\
\end{split}
\end{equation}

\begin{figure}[t]
\begin{center}
\includegraphics[width=7cm]{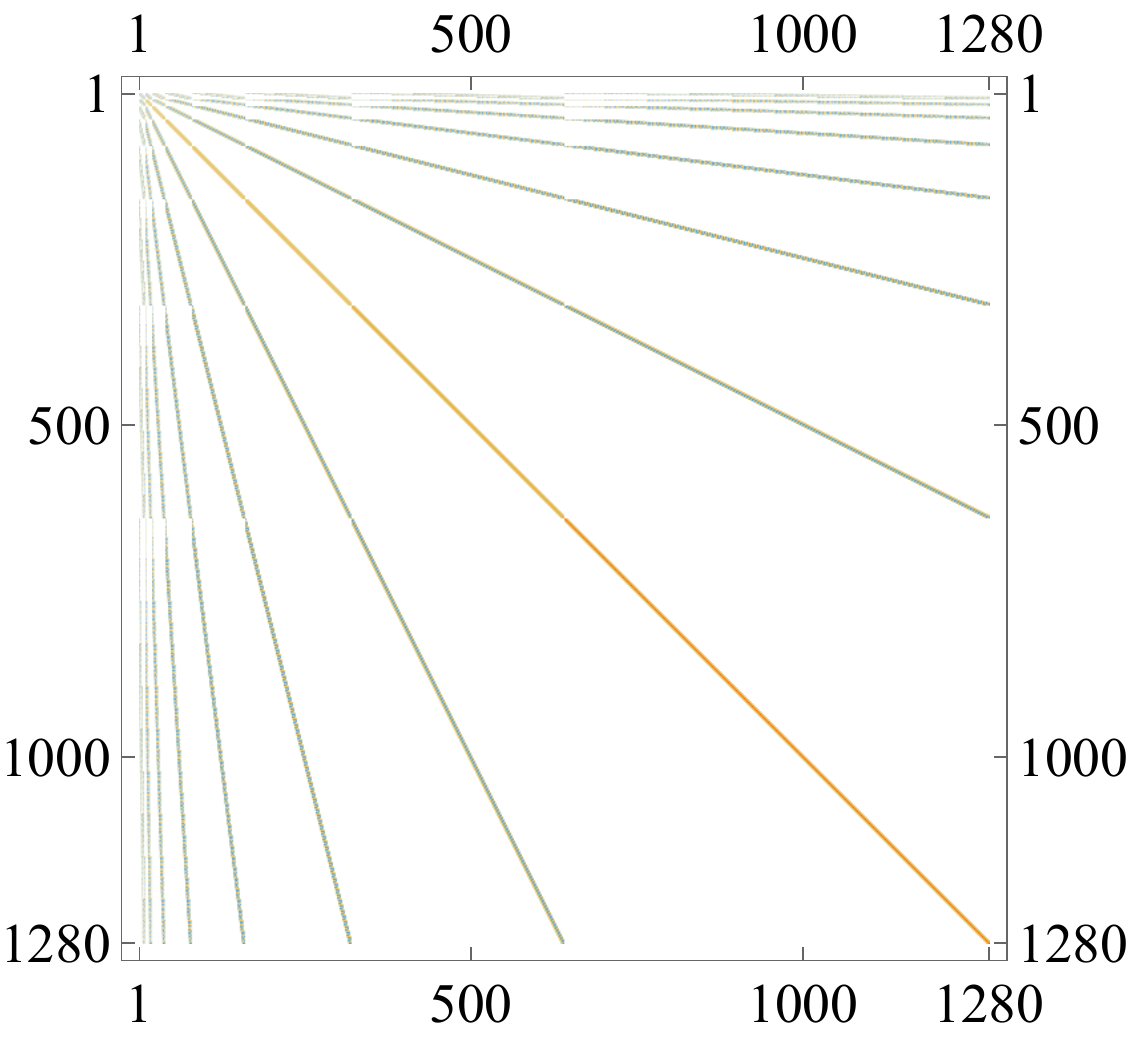}
\label{fig:2}
\end{center}
\caption{Visualization of the $1280 \times 1280$ coupling matrix $K$ in the wavelet basis for a one dimensional free scalar field.  The system size is $L=10$ and the maximum scale is ${\lmax}=6$ so that the total number of modes is $V=1280$.  The diagonal stripe indicates couplings within a given scale while the off diagonal stripes represent couplings between scales.}
\label{fig:Kmatrix}
\end{figure}

The values of these coupling overlap integrals for $K=3$ are obtained from the following relations.  First we use the scaling function components defined in Eq. \ref{Father}
\begin{equation}
\begin{array}{lll}
h_0&=&\frac{1}{16\sqrt{2}}(1+\sqrt{10}+\sqrt{5+2\sqrt{10}})\\
h_1&=&\frac{1}{16\sqrt{2}}(5+\sqrt{10}+3\sqrt{5+2\sqrt{10}})\\
h_2&=&\frac{1}{16\sqrt{2}}(10-2\sqrt{10}+2\sqrt{5+2\sqrt{10}})\\
h_3&=&\frac{1}{16\sqrt{2}}(10-2\sqrt{10}-2\sqrt{5+2\sqrt{10}})\\
h_4&=&\frac{1}{16\sqrt{2}}(5+\sqrt{10}-3\sqrt{5+2\sqrt{10}})\\
h_5&=&\frac{1}{16\sqrt{2}}(1+\sqrt{10}-\sqrt{5+2\sqrt{10}})\\
\end{array}
\end{equation}
The coefficients $g_n=(-1)^n h_{5-n}$.  The coefficients $D^0_{m,n}=D^0_{n,m}$ with
\begin{equation}
\begin{split}
D^0_{0,0}&=5.2576013450,\\
D^0_{0,1}&=-3.3828986455 \\
D^0_{0,2}&=0.87333354692,\\
D^0_{0,3}&=-0.11139112377\\
D^0_{0,4}&=-5.3243362257\times 10^{-3},
\end{split}
\end{equation}
and $D^0_{m,n}=0$ for $|m-n|>4$.  Because the derivatives of translations of the father functions form a partition of unity~\cite{BP2013}
\[
\sum_n n \partial_xs^{\lmin}_n(x)=1,
\]
the coefficients satisfy the following constraint
\[
\sum_n nD^0_{m,n}=0.
\]

The other coefficients are
\begin{equation}
\begin{array}{lll}
D^{0,l}_{a,b}&=&2^{2(l+1)}\bra{a} [H(l)]^{l+1} D(l) G^T(l)\ket{b}\\
D^{l,j}_{a,b}&=&2^{2(l+1)}\bra{a} G(l,j)[H(l,j)]^{l-j} D(l,j) G^T(l,j)\ket{b}
\end{array}
\end{equation}
where the scale dependent matrices are 
\begin{equation}
\begin{array}{lll}
H(l)&=&\sum_{m,n=0}^{2^{(l+1)}(L+4)-5}h_{n-2m}\ket{m}\bra{n} \\
H(l,j)&=&\sum_{m,n=0}^{2^{(l-j)}(2 (2^j L-4))-5}h_{n-2m}\ket{m}\bra{n} \\
D(l)&=&\sum_{m,n=0}^{2^{(l+1)}(L+4)-5}D^0_{m,n}\ket{m}\bra{n} \\
D(l,j)&=&\sum_{m,n=0}^{2^{(l-j)}(2 (2^j L-4)-5}D^0_{m,n}\ket{m}\bra{n} \\
G(l)&=&\sum_{m,n=0}^{2^{(l+1)}(L+4)-5}g_{n-2m}\ket{m}\bra{n} \\
G(l,j)&=&\sum_{m,n=0}^{2^{(l-j)}(2 (2^j L-4)-5}g_{n-2m}\ket{m}\bra{n}. \\
\end{array}
\end{equation}

An example of a $K$ matrix is plotted in Fig. \ref{fig:Kmatrix}. Because the wavelets have compact support, the coupling matrix is sparse having $\simeq 10V\log(V)$ non zero elements, with the factor of $10$ arising from the fact that Daubechies $\mathcal{K}$ wavelets have overlap with $2(2\mathcal{K}-1)$ translates within any given scale.

\subsection{Constructing the ground state of the free field theory}

We would like to encode the vacuum state $\ket{G}$ in Eq. \ref{waveletvac} into a qubit register.  As described in Appendix \ref{discretizedmethod}, let the values $r_j$ be discretized via an $m$ bit string $x_j=x_{j,0}x_{j,1}\ldots x_{j,m-1}$ according to $r_j(x_j)=\delta_{\Phi}(-1)^{x_{j,0}} \sum_{r=1}^{k-1}2^{x_{j,r}}$ with $\delta_{\Phi}$ the field amplitude resolution.  As described in~\cite{JLP2012} (see Sec. \ref{Sec4}), the field resolution scales like $\delta_{\Phi}=O(\sqrt{\frac{\epsilon}{\Vwave E}})$, where $E$ is a bound on the expectation value of the energy during the simulation and $\epsilon$ quantifies the distance between the truncated many body state and the true ground state of the theory.

 The ground state is then represented as a state of $m\times V$ qubits:  
\begin{equation}
\begin{split}
\ket{G}&\approx\mathcal{N}^{-1}\sum_{\{x_{j,r}\in\{0,1\}\}_{j=0,r=0}^{V-1,m-1}}e^{-\frac{1}{2}\vec{r}(\{x_j\})^T K^{1/2} \vec{r}(\{x_j\})}\\
&\ket{x_{0,0}\ldots x_{0,m-1}}_{s^0_{0}}\ldots \ket{x_{V-1,0}\ldots x_{V-1,m-1}}_{w^{{\lmax}}_{L2^{{\lmax}}-1}}
\end{split}
\end{equation}
To construct this ground state using quantum gates one can use the Kitaev-Webb circuit~\cite{KW}.  The cost of that construction is dominated by the $O(V^{2.376})$ time complexity of the classical computation of a matrix decomposition of $K$.

The field operators expressed in the qubit basis are
\begin{equation}
\begin{array}{lll}
\hat{\Phi}^{\scalezero}(n)&=&\delta_{\Phi}\sigma^z_{n,0} \sum_{v=1}^{m-1} 2^{v}(\ket{1}\bra{1})_{n,v}\\
\hat{\Phi}^{\waver{j}}(n)&=&\delta_{\Phi}\sigma^z_{L2^j+n,0} \sum_{v=1}^{m-1} 2^{v}(\ket{1}\bra{1})_{L2^j+n,v}\\
\end{array}
\end{equation}
The momentum operators are not diagonal in the qubit basis so we need to first transform the state to a basis which is diagonal via the $m-1$ qubit quantum Fourier transform (QFT) $\mathcal{F}$ (which acts on all but the sign bit)
\begin{equation}
\begin{array}{lll}
\hat{\Pi}^{\scalezero}(n)&=&\delta_{\Phi}\sigma^z_{n,0}\mathcal{F}^{\dagger} [\sum_{v=1}^{m-1} 2^{v}(\ket{1}\bra{1})_{n,v}]\mathcal{F}\\
\hat{\Pi}^{\waver{j}}(n)&=&\delta_{\Phi}\sigma^z_{L2^j+n,0}\mathcal{F}^{\dagger} [\sum_{v=1}^{m-1} 2^{v}(\ket{1}\bra{1})_{L2^j+n,v}]\mathcal{F}.\\
\end{array}
\end{equation}
%


For completeness, in Appendix \ref{bosenetwork} we also describe how to construct the ground state using an encoding with a bosonic network. Because the state is Gaussian, the preparation proceedure requires only Gaussian operations on single modes or pairs of modes.

\subsection{Particle creation in the free field theory}
\label{momentum}    
Let us consider the steps in a quantum algorithm to create a particle excitation above the vacuum ground state of the free field theory. A simple choice here is to choose the particle's wave function to be the wavelet $\psi(x)=w^r_n(x)$. That is we want to construct the state
\begin{equation}
\hat{b}^{r\dagger}(n)\ket{G}.
\label{excstate}
\end{equation}

The momentum operator in the wavelet basis is~\cite{BP2013} 
\begin{equation}
\begin{array}{lll}
\hat{p}&=&-\Big(\sum_{m,n}:\hat{\Pi}^{\scale}(m)P^k_{m,n}\hat{\Phi}^{\scale}(n):\\
&&+\sum_{(m,l),(n,j)}:\hat{\Pi}^{\wave}(m)P^{l,j}_{m,n}\hat{\Phi}^j(n):\\
&&+\sum_{m,l,n}:\hat{\Pi}^{\wave}(m)P^{l}_{m,n}\hat{\Phi}^{\scale}(n):\Big)
\end{array}
\end{equation}
where
\begin{equation}
\begin{array}{lll}
P^k_{m,n}&=&\int dx s^{\lmin}_m(x)\partial_x s^{\lmin}_n(x)\\
P^{l,j}_{m,n}&=&\int dx w^l_{m}(x)\partial_x w^j_{n}(x)\\
P^{l}_{m,n}&=&\int dx (w^l_{m}(x)\partial_x s^{\lmin}_n(x)+s^{\lmin}_n(x)\partial_x w^l_{m}(x))
\end{array}
\end{equation}
Note that translational shifts in the wave function do not change the momentum.  Furthermore, from the scaling properties of the scaling functions and wavelets, $P^{k}_{m,n}=-P^{k}_{n,m}=2^k P^0_{0,n-m}$, and $P^{l,l}_{m,n}=-P^{l,l}_{n,m}=2^l P^{0,0}_{0,n-m}$.

\begin{table}[ht]
\centering 
\begin{tabular}{c | c | c } 
\hline\hline 
m & $P^{0}_{0,m}$ & $P^{0,0}_{0,m}$ \\ [0.5ex] 
\hline
0 & 0 & 0 \\
1 & 0.745203 & $-$1.32599 \\
2 & $-$0.145203 & 0.146573 \\ 
3 & 0.014612 & $-$0.014612 \\
4 & 0.000342 & $-$0.000342 \\
[1ex]
\hline 
\end{tabular}
\caption{
Values of overlap integrals used to determine the momentum of excited state wave packets for Daubechies $\mathcal{K}=3$ wavelets.  Note $P^{0}_{0,m}=-P^{0}_{0,-m}$ and $P^{0,0}_{0,m}=-P^{0,0}_{0,-m}$, and for $|m|>4$ the values are zero.
\label{table:gsenergy} 
}
\end{table}

For the excited state in Eq. \ref{excstate}, the expectation value of the momentum  (assuming $r\geq k$) is:
\begin{equation}
\bra{G}\hat{b}^r(n) \hat{p} \hat{b}^{r\dagger}(n)\ket{G}=-\frac{i}{2}2^rP^{0,0}_{0,0}=0.
\end{equation}
Finite momentum excited states can be created from a superposition of wavelets.  Consider the state
\begin{equation}
\ket{E}=(\alpha_{r,n} \hat{b}^{r\dagger}(n)+\beta_{r,m} \hat{b}^{r\dagger}(m))\ket{G},
\label{sprexc}
\end{equation}
with $r\geq k$ and $|\alpha_{r,n}|^2+|\beta_{r,m}|^2=1$.  We find 
\begin{equation}
\bra{E}\hat{p}\ket{E}=2^{r} P^{0,0}_{0,n-m}\Im[\alpha_{r,n}\beta_{r,m}^*].\nonumber
\end{equation}
For a given scale, the maximum magnitude momentum eigenstate is obtained for $n-m=-1$, $\alpha_{r,n}=1/\sqrt{2}$, $\beta_{r,m}=\mp i/\sqrt{2}$ in which case $\bra{E}\hat{p}\ket{E}=\pm 2^{r-1}P^{0,0}_{0,-1}=\pm 0.663\times 2^{r}$.  
  Hence the maximum momentum of a single particle state is 
  \begin{equation}
  p_{\rm max}=0.663\times 2^{{\lmax}}.
  \label{maxmom}
  \end{equation}

Define a generalised single particle excitation $f^{\dagger}\ket{G}$ where $\hat{f}^{\dagger}=\sum_{r,n}\alpha_{r,n}b^{r \dagger}(n)$ and $\sum_{r,n}|\alpha_{r,n}|^2=1$.  We follow the approach in Ref.~\cite{JLP2012} and introduce an ancillary qubit $a$ interacting with  the register qubits via
\begin{equation}
\hat{H}_{\psi}=\hat{f}^{\dagger}\otimes (\ket{1}\bra{0})_a+\hat{f}\otimes (\ket{0}\bra{1})_a.
\end{equation}
If we can simulate the evolution by $\hat{H}_{\psi}$, then $e^{-i\hat{H}_{\psi}\pi/2}\ket{G}\ket{0}_a=-i\hat{f}^{\dagger}\ket{G}\ket{1}_a$ and we have the excited state up to a phase with no entanglement left between the ancilla and the register.  The Hamiltonian written out explicitly in the qubit representation is
\begin{equation}
\begin{split}
\hat{H}_{\psi}&=\frac{\delta_{\Phi}}{\sqrt{2}}\sum_{r,n}\Bigg[\Big(\Re[\alpha_{r,n}]\sqrt{\gamma^{[\vec{w}]}(r)}
\sigma^z_{L2^r+n,0} [\sum_{v=1}^{m-1} 2^{v}(\ket{1}\bra{1})_{L2^r+n,v}]\\
&+\Im[\alpha_{r,n}]\frac{1}{\sqrt{\gamma^{[\vec{w}]}(r)}}\sigma^z_{L2^r+n,0}\mathcal{F}^{\dagger} [\sum_{v=1}^{m-1} 2^{v}(\ket{1}\bra{1})_{L2^r+n,v}]\mathcal{F}\Big)\otimes\sigma^x_a\\
&
+\Big(\Im[\alpha_{r,n}]\sqrt{\gamma^{[\vec{w}]}(r)}
\sigma^z_{r,0} [\sum_{v=1}^{m-1} 2^{v}(\ket{1}\bra{1})_{r,v}]\\
&-\Re[\alpha_{r,n}]\frac{1}{\sqrt{\gamma^{[\vec{w}]}(r)}}\sigma^z_{r,0}\mathcal{F}^{\dagger} [\sum_{v=1}^{m-1} 2^{v}(\ket{1}\bra{1})_{r,v}]\mathcal{F}\Big)\otimes\sigma^y_a\Bigg],
\end{split}
\end{equation}
The evolution generated by these non commuting terms can then be simulated efficiently by Trotter decomposition~\cite{BACS05}.  Note that the overhead cost to implement the QFT is $O(m^2)$.  

An method to prepare single particle excitation in a bosonic network encoding is given in Appendix \ref{bosenetwork}.

\subsection{Interacting field theory}
The Hamiltonian including interactions in the wavelet basis is
\begin{equation}
\hat{H}=\hat{H}(a)+\hat{H}(b)+\hat{H}(ab)+\hat{H}^{(I)},
\end{equation}
where the interaction term is:
\begin{equation}
\begin{split}
\hat{H}^{(I)}=\frac{\lambda_0}{4!}\sum_{z's\in\{w,s\}}\sum_{j's}\sum_{n's}  \int dx\ f^{z_1,j_1}_{n_1}(x)f^{z_2,j_2}_{n_2}(x)f^{z_3,j_3}_{n_3}(x)f^{z_4,j_4}_{n_4}(x)\\
:\hat{\Phi}^{[z_1]j_1}(n_1)\hat{\Phi}^{[z_2]j_2}(n_2)\hat{\Phi}^{[z_3]j_3}(n_3)\hat{\Phi}^{[z_4]j_4}(n_4):,
\end{split}
\end{equation}
where
\begin{equation}
f^{z,j}_{n}=\left\{\begin{split} s^{\lmin}_n(x)\quad j&=\lmin\ {\rm and}\ z=s \\w^j_{n}(x)\quad j&\geq \lmin\ {\rm and}\ z=w
\end{split}\right.
\end{equation}
Because, the scale functions and wavelets have compact support, the number of non zero summands in the interaction scales like $O(V\log(V))$. 

\section{Resource scaling of the quantum wavelet simulation}
\label{Sec4}
 The overall efficiency of the bosonic field theory simulator in the wavelet basis can be obtained by comparing it with the discretized position basis algorithm which was carefully analysed in Ref.~\cite{JLP2012}.
 In the latter algorithm the real valued field, which is a function of the continuous position degree of freedom, is discretised by treating the volume as finite and composed of $N\in\mathbb{N}$ points equally spaced by physical length $\adis$ in each dimension. The longest wavelength physics that can be captured is $N\adis$ and the highest momentum that can be represented is $1/\adis$. Furthermore, the amplitude for the field at each of the $\Vdis=N^d$ discrete points in space is discretised to values in the set $\delta_{\Phi} [-2^{b-1},\ldots, 2^{b-1}]$ where $b=\log(\Phi_{\tiny \mbox{max}}/\delta_{\Phi})$\footnote{In~\cite{JLP2012} the lowercase position and momentum densities are used which satisfy the canonical commutation relations $[\hat{\phi}(\bm{x}),\hat{\pi}(\bm{y})]=i\adis^{-d}\delta(\bm{x},\bm{y})$. In the main text we normalized the field and momentum densities so that $[\hat{\Phi}(\bm{x}),\hat{\Pi}(\bm{y})]=i\delta(\bm{x},\bm{y})$.}.
 The efficiency of the algorithm is quantified in terms of two important quantities: the total energy bound $E$ such that the evolved state satisfies $\bra{\psi(t)}\hat{H} \ket{\psi(t)}\leq E$ for all times in the simulation, and the error $\epsilon$ which is defined in terms of fidelity of the truncated and discretized many body state with the true state:  $|\bra{\Psi}\Psi_{\rm cut}\rangle|\geq 1-\epsilon$. A cutoff in the maximum field amplitude $\Phi_{\tiny \mbox{max}}=O(\sqrt{\frac{\Vdis E}{m_0^2\epsilon}})$ ensures the above fidelity. By the Fourier relation between conjugate variables, the momentum cutoff is $\Pi_{\tiny \mbox{max}}=\delta_{\Phi}^{-1}$, and upper bounding the expectation values of $\hat{\Pi}(\bm{x})$ and $\hat{\Pi}^2(\bm{x})$ in terms of energy, it suffices to choose $\Pi_{\tiny \mbox{max}}=O(\sqrt{\frac{\Vdis E}{\epsilon}})$. The number of qubits needed for the simulation is then $n=\Vdis b=O(\Vdis\log(\frac{\Vdis E}{m_0\epsilon}))$. In the massive case, two point correlators decay exponentially with separation, and $\Vdis$ need only scale logarithmically with $\epsilon$. 
 
 The asymptotic scaling for the number of quantum gates needed to simulate particle scattering is found by suming the gates for the following steps:  free field ground state preparation, excited state preparation by adiabatic turn on of particle creation interaction, adiabatic turn on of interaction terms in the Hamiltonian, and finally measurement of scattering probabilities. It is shown that the total number of gates is a small polynomial in $1/\epsilon$ in the weak coupling regime, and in the strong coupling regime there is an additional overhead of a polynomial in the momentum $p$ of the colliding particles, the number of outgoing particles, and the distance from the phase transition such that the overall scaling for a simulation of duration $t$ is $O(p^{d+1+o(1)}(t\Vdis)^{1+o(1)})$. 

In the wavelet basis, for the one dimensional case $d=1$, the number of modes is $\Vwave=L2^{{\lmax}+1}$ (Eq. \ref{nummodes}) and for arbitrary dimension, $\Vwave=(L2^{{\lmax}+1})^d$. The longest wavelength physics that can be captured is $L\awave$ and the highest momentum scale, from Eq. \ref{maxmom}, is $2^{l_{\rm max}}\awave$. In order to compare the resource scaling with the case of discrete basis we need to equate the longest wavelength and highest momentum scales in the two descriptions, namely,
\begin{equation}
\begin{split}
N\adis &= L\awave,\\
\frac{1}{\adis} &= \frac{2^{\lmax}}{\awave},
\end{split}
\end{equation}
which implies
\begin{equation}
 N = L2^{\lmax} \Rightarrow \Vwave = 2^d~\Vdis.
 \end{equation}
In dimensions $d=1,2$ or $3$ the number of modes used in both simulations are very similar.
The same arguments that led to the scaling of the maximum field amplitude $\Phi_{\tiny \mbox{max}}$ apply, namely we are truncating a field on $\Vwave$ modes by cutoffs in the field amplitude at $\Phi_{\tiny \mbox{max}}$. Hence the scalings of $\Phi_{\tiny \mbox{max}},\Pi_{\tiny \mbox{max}}$ and the total number of qubits $b$ is the same as in the discretised position basis where $\Vdis$ is replaced by $\Vwave$.

The number of quantum gates to perform a quantum simulation incurs only a penalty of replacing $V$ with $V\log(V)$ in the scaling formulae relative to the discretized basis encoding.  The reason is that in the wavelet basis the terms in the free field and interacting Hamiltonians couple across all scales as opposed to the discretized position basis where only nearest neighbour modes are coupled. 
Because the wavelets have compact support, the number of summands in $\hat{H}$ scales like $O(\Vwave\log(\Vwave))$. The first step of constructing the ground state of the free field Hamiltonian has time cost $O(\Vwave^{2.376})$, the same form as in the discretised bases, which is obtained from the worse case scaling assuming a dense correlation matrix.  During particle creation and simulated evolution steps, the aforementioned additional terms in the Hamiltonian using the wavelet basis means scaling with respect to $V$ in the discretised basis should be replaced by $\Vwave\log(\Vwave)$. 
Finally measurement has the same scaling in either basis. A notable advantage of using the wavelet basis is that particle creation and measurement can be done at a variety of different length/energy scales without further transformations on the system.

\section{Entanglement Entropy}
\label{Sec3}
In order to illustrate of the utility of the wavelet representation we turn to the calculation of entropic quantities. The entanglement entropy for the free scalar bosonic field theory has been calculated for $d\geq 1$~\cite{Hertzberg}.  In addition the interacting case has be treated perturbatively for $\hat{\Phi}^4$ theory in $d=3$.~\cite{Hertzberg}, with the main result being that the bare mass $m_0$ is replaced by the renormalised mass $m_r$ at the renormalisation scale of zero momentum.
We focus on the $d=1$ case here. For a system of size $L$ and subregion $A$ of length $\ell$, the entanglement entropy
\begin{equation}
S(\rho_A)=-\tr [\rho\log(\rho_A)]
\end{equation}
of the free scalar bosonic theory was calculated by Calabrese and Cardy~\cite{Cardy}. In the massive case we have
\begin{equation}\label{eq:entmassive}
\begin{array}{lll}
S_{A}&=&-\frac{1}{12}\log(m_0^2a^2)\\
&=&\frac{1}{6}\log(\xi/a)
\end{array}
\end{equation}
where $\xi=m_0^{-1}$ is the correlation length. For the massless case, which corresponds to the 1+1 dimensional bosonic conformal field theory (CFT) with central charge $c=1$,
\begin{equation}\label{eq:entmassless}
S_{A}\approx \left\{\begin{array}{cc}   \frac{1}{3}\log\Big(\frac{L}{\pi a}\sin(\pi \ell/L)\Big)+C_{\rm per} & {\rm periodic\ boundaries} \\ \frac{1}{6}\log\Big(\frac{2L}{\pi a}\sin(\pi \ell/L)\Big) +C_{\rm open}& {\rm open\ boundaries}\end{array}\right.
\end{equation}
where $C_{\rm per},C_{\rm open}$ are constant correction terms.
\begin{figure}[t]
\begin{center}
\includegraphics[width=\columnwidth]{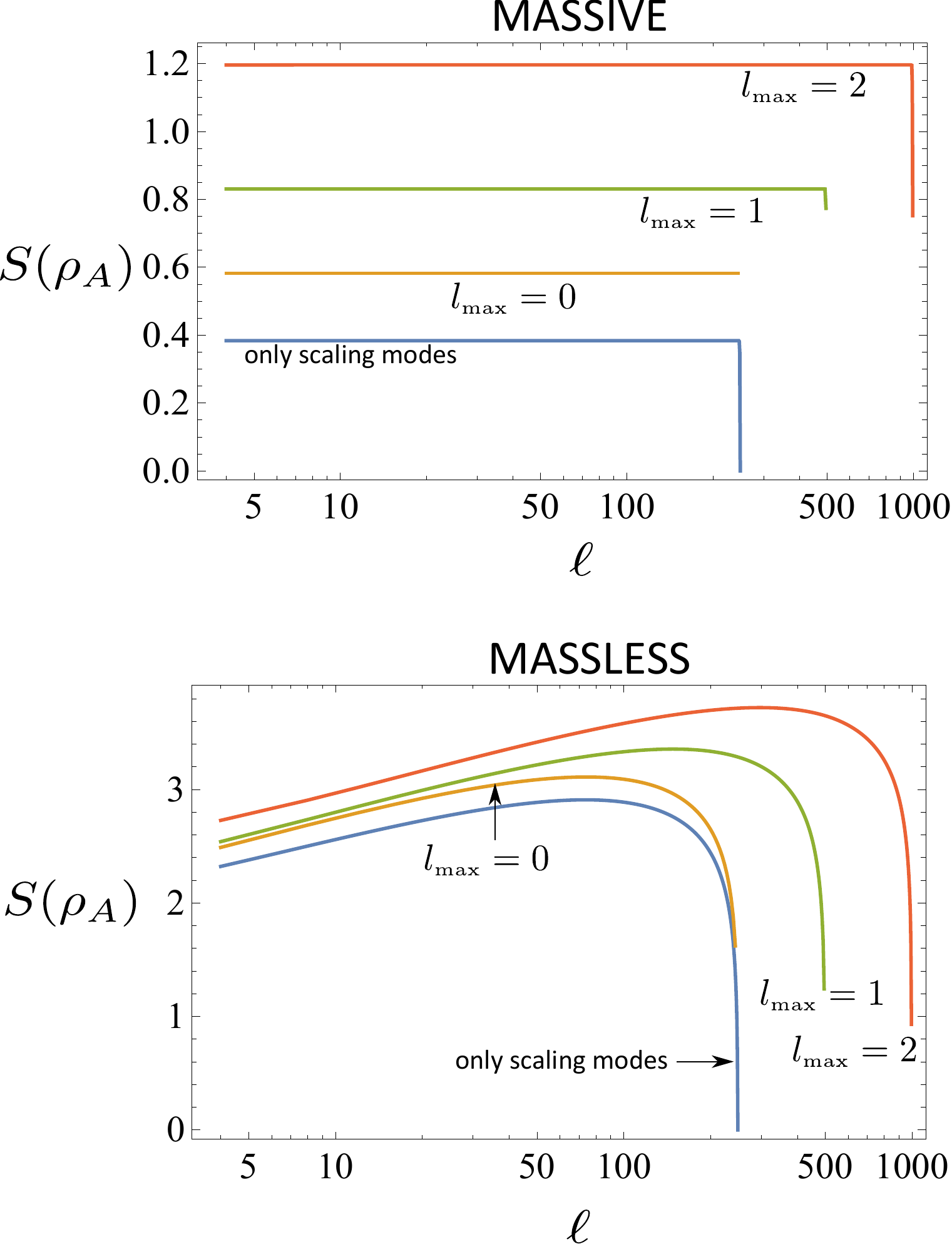}
\label{fig:entropy}
\end{center}
\caption{Plots of subsystem entropy for a $d=1$ free scalar field theory represented in the wavelet basis.  The subsystem $A$ corresponds to a contiguous block of size $\ell$.  Entropies were computed for ground states of Hamiltonians with ${\lmax}=0,1,2$, as well as for the fully renormalized Hamiltonian keeping only the scaling modes.  (a) Massive case $(m_0=1)$ (b) Massless case $(m_0=0)$ plotted on a log scale.  For small $\ell\ll L$, the entropy is linear in $\log(\ell)$ and the central charge can be extracted from the portionality constant.}
\end{figure}

\subsection{Calculating entanglement entropy in the wavelet basis}
%

Recall that the covariance matrix $\Gamma$ of a Gaussian state is defined as
\beq
\Gamma_{j,k} = \Re[ \tr[\rho (\hat{\vec{r}}_j-\langle \hat{\vec{r}}_j\rangle) (\hat{\vec{r}}_k-\langle \hat{\vec{r}}_k\rangle)]],
\eeq
where $\langle\hat{\vec{r}}_j\rangle$ is the expectation value of $j$-th element of the vector $\hat{\vec{r}} = (\hat{q}_1, ..., \hat{q}_V, \hat{p}_1, ..., \hat{p}_V)^T$ of quadrature operators on the $N$ modes. The information contained in the covariance matrix completely determines the entanglement properties of a Gaussian state. Explicit calculations for Gaussian states are performed making use of the \textit{symplectic spectrum} of $\Gamma$. Let us introduce the symplectic form $\Omega$,
\beq
\Omega_{j,k} = -i [ \hat{\vec{r}}_i, \hat{\vec{r}}_j ] = \left(\begin{array}{cc}\bm{0} & \bm{1}_V \\-\bm{1}_V & \bm{0}\end{array}\right) \, ,
\eeq
which is a skew-symmetric matrix that incapsulates the canonical commutation relations of the quadrature operators. For a Gaussian state $\rho$ with covariance matrix $\Gamma$, the positive elements of the $V$ pairs of eigenvalues $\{ \pm \sigma_i \}$ of the matrix product $i \Gamma \Omega$ are called \textit{symplectic eigenvalues}. The entropy for Gaussian subsystem $\rho_A$ corresponding to $N_A$ modes is
\begin{align}
\label{formulasetext}
S(\rho_A) = \sum_{\mathclap{\{ \sigma^A_j \}}}  \bigl[   ( \sigma^A_j + \tfrac{1}{2} ) \log_2   ( \sigma^A_j + \tfrac{1}{2} ) -  ( \sigma^A_j - \tfrac{1}{2} ) \log_2  ( \sigma^A_j - \tfrac{1}{2}  ) \bigr],
\end{align}
calculated using the reduced symplectic spectrum $\{\sigma^A_1, \dotsc, \sigma^A_{N_A}\}$ obtained deleting the rows and columns corresponding to the complementary modes from the covariance matrix. 

\begin{figure}[t]
\begin{center}
\includegraphics[width=\columnwidth]{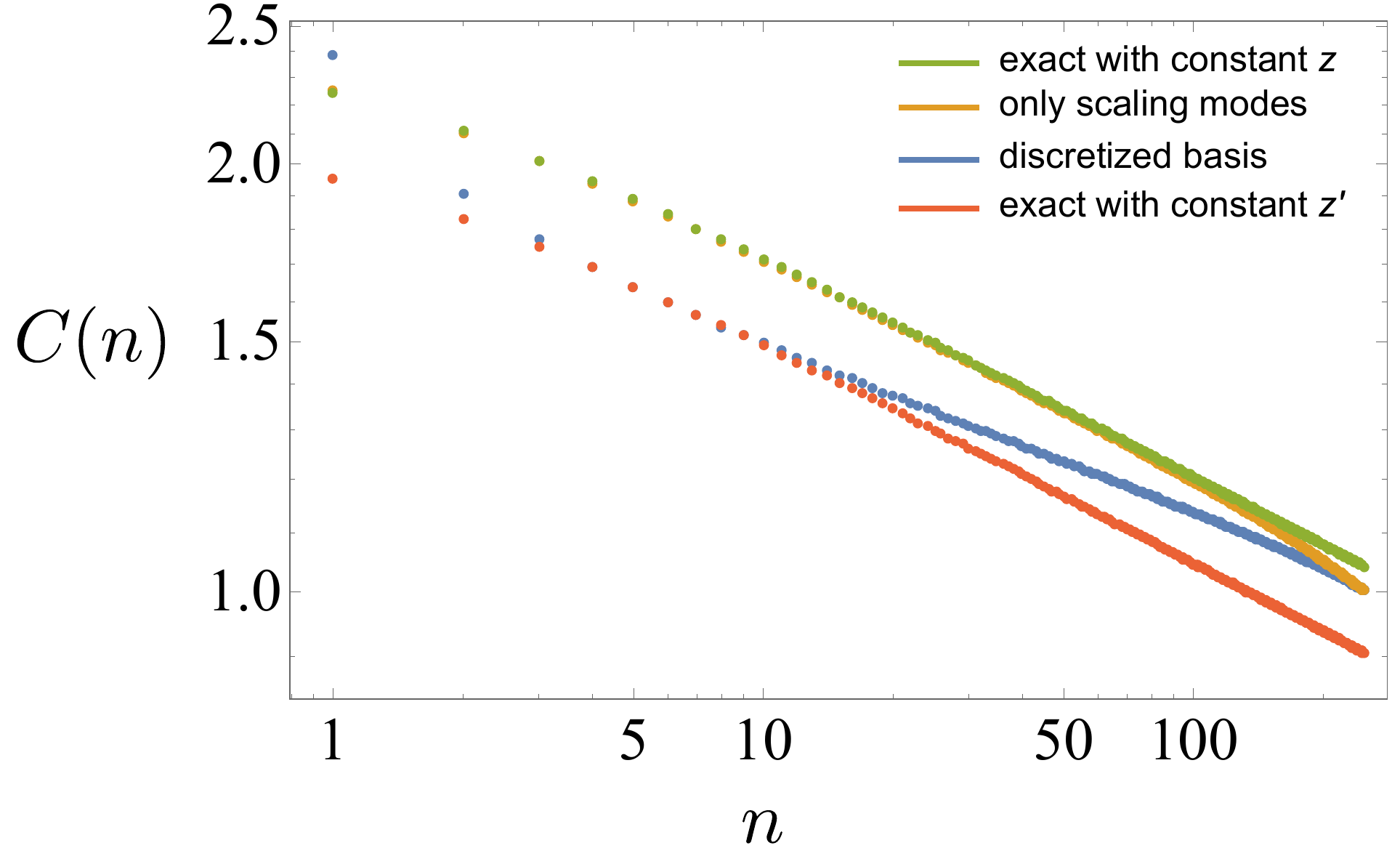}
\label{fig:scalecorr}
\end{center}
\caption{Two point field-field vacuum correlations for the massless free field. The plots are in two bases: the wavelet basis keeping only the $L=500$ scaling modes in the description, and the discrete position basis (see Appendix \ref{discretizedmethod}) consisting of $N=500$ modes. Also shown is the exact correlation function for a free scalar boson CFT $C(n) = -\frac{1}{4\pi}\mbox{ln}(n^2) + \mbox{const}$, plotted for two choices of the constant $z=0.92$ and $z'=0.78$ that best fit the computed correlations in the two bases.}
\end{figure}

The same prescription also applies to computing entanglement entropy in the wavelet basis where the covariance matrix is constructed using the coupling matrix in Eq.\ref{Kmatrix}:
\begin{equation}
\Gamma=\frac{1}{2}\left(\begin{array}{cc}K^{-1/2} & 0 \\0 & K^{1/2}\end{array}\right).
\label{corrmat}
\end{equation}
We computed how the subregion entanglement entropy scales with the physical size $\ell$ of the subregion for both the massive and the massless cases. Fig. \ref{fig:entropy} shows this scaling for a system with total size $L=500$ and for $l_{\tiny \mbox{max}}=0,1,2$ which corresponds to keeping smaller and smaller scale features in the description. For $l_{\tiny \mbox{max}}=0$ the entanglement entropy of the subsystem receives contributions from scaling modes at the base scale $l=0$ and wavelet modes only at length scales $l=0$, which have support in the interval of physical size $\ell$. For $l_{\tiny \mbox{max}}=1,2$ the entanglement entropy also receives contributions from the wavelet modes at $l=1$ and $l=1,2$ respectively. Indeed, we reproduce the expected scaling of entanglement entropy in both the massive and massless cases, Eq. \ref{eq:entmassive} and Eq. \ref{eq:entmassless} respectively. We also plot the entanglement entropy obtained from the (renormalized) description of the ground state by only keeping the scaling modes (i.e. only keeping the $K_{ss}$ block in the coupling matrix, Eq.\ref{Kmatrix}). In the massive case, because correlations fall off exponentially, one can take $L$ to be $O(1)$ and still capture the essential physics.

In the massless case, we can estimate the central charge from the slope of the linear part of the plot in the regime $\ell\ll L$. We find $c \approx 1.004$ which agrees with the central charge $c=1$ of the scalar bosonic CFT. Interestingly, an accurate value is obtained by only keeping the scaling modes. We also plot in Fig. \ref{fig:scalecorr} the two point field-field correlations, once again only keeping the scaling field modes in the description, which agree with the exact field-field correlation scaling for a free scalar boson CFT,
\begin{equation}
C(n) = -\frac{1}{4\pi}\mbox{ln}(n^2) + \mbox{const}.
\end{equation}
As shown in Fig. \ref{fig:scalecorr} the correlations in the scale mode degrees of freedom $C(n) = \langle\hat{\Phi}^{\scaler{\lmin}}(L/2),\hat{\Phi}^{\scaler{\lmin}}(n)\rangle$ ($n=1,2,\ldots L/2-1$) are in fact a better fit to the CFT prediction than the correlations in a discretised position basis. 
These results suggest that the scaling modes accurately capture the large scale properties of the system, and indeed are the basis for the description of the system at the renormalization fixed point.

\section{Conclusions}
We have shown that scalar bosonic quantum field theories can be simulated efficiently on a quantum computer using a wavelet basis. Without compromising overall efficiency, our algorithm reorganizes the quantum state and its evolution into sectors at different length scales. We anticipate this could be useful to study renormalization flow and is a natural setting for characterizing fields in terms of finite bandwidth detectors. We computed the entanglement entropy in a $d=1$ free field theory and found that the wavelet basis conveniently divides the state into long range entangled and short range entangled degrees of freedom at the massless critical point. The correlations and entanglement of the bosonic CFT are simply computed from the largest scale degrees of freedom indicating that the coarse scale Hamiltonian is the renormalised Hamiltonian for the system at criticality. 

For future work we note that an efficient quantum algorithm is known~\cite{Hoyer1997,FW1999, Arguello2009} for performing Daubechies$\mathcal{K}$-wavelet transforms on an $m$ qubit register in $O(m^2)$ gates \footnote{An approximate transform which limits gate precision to $2^{-r}$ has complexity $O(rm)$~\cite{FW1999}}. This algorithm translates between the discretized position space representation of a single particle and the wavelet representation. It would be of interest to adapt this to quantum simulations of multiparticle strongly correlated systems.

Finally, we remark that there could be interesting connections between the multi-scale representation of quantum many-body states using the MERA and the wavelet basis described here. In the wavelet basis, the wavelet modes capture the short-range entangement at any given length scale, while in the MERA the same role is played by local disentangling and coarse-graining transformations.

\acknowledgements
We thank Dominic Berry and Alexei Gilchrist for helpful discussions. BCS acknowledges financial support from AITF and NSERC.

\appendix

\section{Preparing the ground state using a discretised position basis}
\label{discretizedmethod}
In this appendix we review how the state representing the quantum field is encoded in Ref.~\cite{JLP2012}.
The Hamiltonian $\hat{H}^{(0)}$ can be obtained as the continuum limit of a discrete system with $\Vdis$ bosons with canonical position and momenta variables $\{\hat{q}_j\}$ and $\{\hat{p}_j\}$ respectively, which satisfy $[\hat{q}_j,\hat{p}_k]=i\delta_{j,k}$.  Consider the following Hamiltonian for bosons on the sites of a cubic lattice of size $\Vdis=L^d$ with a uniform lattice spacing $a$ 
\begin{equation}
\hat{H}^{(0)}=\sum_{m=1}^{\Vdis} \frac{\hat{p}_m^2}{2\mu}+\frac{\lambda}{2}\sum_m \hat{q}^2_m+\kappa \sum_{\langle m,n\rangle}(\hat{q}_m-\hat{q}_n)^2,
\label{discrete}
\end{equation}
where the sum over $\langle m,n\rangle$ is over nearest neighbour pairs at positions $\vec{x}_m$ and $\vec{x}_n$. The continuum limit is obtained by taking
\begin{itemize}
\item
$\Vdis\rightarrow \infty$
\item
Positions $\bm{x}_m=\adis \bm{m},\quad \bm{m}\in\mathbb{Z}^d$ 
\item
$\hat{q}_m\rightarrow \hat{\Phi}(\bm{x}_m)$
\item
$\sum_m \rightarrow \frac{1}{\adis^d} \int d^dx$
\item
$(\hat{q}_m-\hat{q}_n)\rightarrow \adis (\vec{\nabla} \hat{\Phi}(\bm{x}))_{\langle m,n\rangle}$
\item
$\kappa=\adis^{d-2}\bar{\kappa},\quad \mu=\adis^d \bar{\mu},\quad \lambda=\adis^d\bar{\lambda}$
\item
Rescaling:  $\hat{\Phi}(\bm{x})\rightarrow  \bar{\kappa}^{-1/2}\hat{\Phi}(\bm{x})$,
\end{itemize}
which leads to the Hamiltonian density
\begin{equation}
\hat{\mathcal{H}}^{(0)}=\frac{1}{2}(\bar{\mu}\bar{\kappa}^{-1}\hat{\Pi}(\bm{x},t)^2+(\vec{\nabla}\hat{\Phi}(\bm{x},t))^2+\bar{\lambda}\bar{\kappa}^{-1}\hat{\Phi}(\bm{x},t)^2).
\end{equation}
Setting $\bar{\mu}\bar{\kappa}^{-1}\rightarrow 1$ and $\bar{\lambda}\bar{\kappa}^{-1}=\lambda \mu^{-1}=m^2_0$ we obtain the Hamiltonian density for the free field interaction in Eq. \ref{Ham}. 

The descretized version of the Hamiltonian (Eq. \ref{discrete}) is compactly written:
\begin{equation}
\hat{H}^{(0)}= \frac{1}{2} \hat{\vec{r}}^T A \hat{\vec{r}},
\end{equation}
where $\hat{\vec{r}}$ is the $2\Vdis$ dimensional vector of position and momenta operators, $\hat{\vec{r}}=(\hat{q}_1,\ldots \hat{q}_\Vdis,\hat{p}_1,\ldots \hat{p}_\Vdis)^T$, and
\begin{equation}
A=\left(\begin{array}{cc}K & 0 \\0 & \bm{1}_\Vdis\end{array}\right),
\end{equation}
where 
\begin{equation}
K_{i,j}=(4d+m^2_0)\delta_{i,j}-2\delta~~~(i\in {\rm neighborhood} j).
\label{discreteK}
\end{equation}

 The covariance matrix 
associated with a state $\rho$ is defined $\Gamma_{j,k}= \Re[ \tr[\rho (\hat{\vec{r}}_j-\langle \hat{\vec{r}}_j\rangle) (\hat{\vec{r}}_k-\langle \hat{\vec{r}}_k\rangle)]]$, where $\langle\hat{\vec{r}}_j\rangle$ is the expectation value of $j$-th element of $\hat{\vec{r}}$ and where $K$ is defined in Eq. \ref{discreteK}.
 The ground state (vacuum) of this system can then be expressed as a Gaussian in the position basis:
\begin{equation}
\ket{G}=\mathcal{N}^{-1}\int_{-\infty}^{\infty} dq_1\ldots \int_{-\infty}^{\infty} dq_\Vdis\ e^{-\frac{1}{2}\textbf{q}^T K^{1/2} \textbf{q}}\ket{q_1}\ldots \ket{q_\Vdis},
\end{equation}
where $\mathcal{N}^{-1}=\det(K^{1/2})^{1/4}/\pi^{\Vdis/4}$ is the normalisation and $\bm{q}=(q_1,\ldots,q_\Vdis)^T$.

 The values of $q_j$ are discretized via a $b$ bit string $x_j=x_{j,0}x_{j,1}\ldots x_{j,b-1}$ according to $q_j(x_j)=\delta_{\Phi} (-1)^{x_{j,0}} \sum_{r=1}^{b-1}2^{x_{j,r}} $. The ground state can be represented as a state of $b\times \Vdis$ qubits:
\begin{equation}
\begin{array}{lll}
\ket{G}&\approx&\mathcal{N}^{-1}\sum_{\{x_{j,r}\in\{0,1\}\}_{j=1,r=0}^{\Vdis,b-1}}e^{-\frac{1}{2}\bm{q}(\{x_j\})^T K^{1/2} \bm{q}(\{x_j\})}\\
&&\ket{x_{0,0}\ldots x_{0,k-1}}\ldots \ket{x_{\Vdis-1,0}\ldots x_{\Vdis-1,k-1}}
\end{array}
\end{equation}

\section{Wavelets for higher dimensions $d>1$} 
\label{higherD}
The wavelet representation for scalar field theories can be straightforwardly generalised to higher dimensions as described in Ref.~\cite{BP2013}. For completeness, we include the argument. In $d=3$, for example, $\bm{n}=(n_x,n_y,n_z)\in\mathbb{Z}^3$ and the scale functions $s^{\lmin}_{\bm{n}}(\bm{x})=s^{\lmin}_{n_1}(x_1)s^{\lmin}_{n_2}(x_2)s^{\lmin}_{n_2}(x_3)$, where $\bm{x}$ is now a position vector in $\mathbb{R}^3$ and $\bm{n}=(n_x,n_y,n_z)\in\mathbb{Z}^3$ becomes a displacement vector. The generalised wavelets $w^m_{\textbf{n},\alpha}(\bm{x})$ are defined by seven different forms (distinguished by the index $\alpha$):
\begin{equation}
\begin{split}
w^m_{\textbf{n},1,k_3}&=s^{\lmin}_{n_1}(x_1)s^{\lmin}_{n_2}(x_2)w^{k_3}_{n_2}(x_3)\quad  m=\max(k,k_3)\\
w^m_{\textbf{n},2,k_3}&=s^{\lmin}_{n_1}(x_1)w^{k_2}_{n_2}(x_2)s^{\lmin}_{n_2}(x_3)\quad  m=\max(k,k_2)\\
w^m_{\textbf{n},3,k_3}&=w^{k_1}_{n_1}(x_1)s^{\lmin}_{n_2}(x_2)s^{\lmin}_{n_2}(x_3)\quad m=\max(k,k_1)\\
w^m_{\textbf{n},4,k_3}&=s^{\lmin}_{n_1}(x_1)w^{k_2}_{n_2}(x_2)w^{k_3}_{n_2}(x_3)\quad m=\max(k,k_2,k_3)\\
w^m_{\textbf{n},5,k_3}&=w^{k_1}_{n_1}(x_1)w^{k_2}_{n_2}(x_2)s^{\lmin}_{n_2}(x_3)\quad m=\max(k,k_1,k_2)\\
w^m_{\textbf{n},6,k_3}&=w^{k_1}_{n_1}(x_1)s^{\lmin}_{n_2}(x_2)w^{k_3}_{n_2}(x_3)\quad m=\max(k,k_1,k_3)\\
w^m_{\textbf{n},7,k_3}&=w^{k_1}_{n_1}(x_1)w^{k_2}_{n_2}(x_2)w^{k_3}_{n_2}(x_3)\quad m=\max(k_1,k_2,k_3)
\end{split}
\end{equation}
The mode operators $\hat{\Phi}$ and $\hat{\Pi}$ are now indexed as:
\begin{equation}
\begin{split}
\hat{\Phi}^{\scale}(n,t) &\rightarrow \hat{\Phi}^{\scale}(\bm{n},t),~~~\hat{\Phi}^{\wave}(n,t) \rightarrow \hat{\Phi}^{\wave}(\bm{n},\alpha,t),\\
\hat{\Pi}^{\scale}(n,t) &\rightarrow \hat{\Pi}^{\scale}(\bm{n},t),~~~\hat{\Pi}^{\wave}(n,t) \rightarrow \hat{\Pi}^{\wave}(\bm{n},\alpha,t),
\end{split}
\end{equation}
where the discrete field operators satisfy the following equal time commutation relations (assuming here that $k\leq r,s$):
\begin{equation}
\begin{split}
\ [\hat{\Phi}^{\scale}(\bm{n}),\hat{\Phi}^{\scale}(\bm{m})]&=0,\quad [\hat{\Pi}^k(\bm{n}),\hat{\Pi}^{\scale}(\bm{m})]=0\\
\ [\hat{\Phi}^{\scale}(\bm{n}),\hat{\Pi}^{\scale}(\bm{m})]&= i \delta_{\bm{n},\bm{m}}\\
\ [\hat{\Phi}^{\waver{r}}(\bm{n},\alpha),\hat{\Phi}^{\waver{s}}(\bm{m},\beta)]&=0\\
\hat{\Pi}^{\waver{r}}(\bm{n},\alpha),\hat{\Pi}^{\waver{s}}(\bm{m},\beta)]&=0\\
\ [\hat{\Phi}^{\waver{r}}(\bm{n},\alpha),\hat{\Pi}^{\waver{s}}(\bm{m},\beta)]&=i\delta_{\alpha,\beta}\delta_{r,s}\delta_{\textbf{n},\textbf{m}}\\
\ [\hat{\Phi}^{\waver{r}}(\bm{n},\alpha),\hat{\Phi}^{\waver{s}}(\bm{m})]&=0,\quad [\hat{\Pi}^{\waver{r}}(\textbf{n},\alpha),\hat{\Pi}^{\waver{s}}(\bm{m})]=0\\
\ [\hat{\Phi}^{\waver{r}}(\bm{n},\alpha),\hat{\Pi}^{\waver{s}}(\bm{m})]&=0,\quad [\hat{\Pi}^{\waver{r}}(\bm{n},\alpha),\hat{\Phi}^{\waver{s}}(\bm{m})]=0
 \end{split}
 \end{equation}
\section{Adaptation of the simulation to bosonic encoding}
\label{bosenetwork}
Rather than discretising the amplitude of the register modes using qubits we could instead opt to directly use $V$ distinguishable bosonic modes with position basis states $\{\ket{q}_{\hat{q}_0},\ldots, \ket{q_{V-1}}_{\hat{q}_{V-1}}\}$.  In this case the mode operators in Eq. \ref{modeops} are just the position operators $\{\hat{q}_j\}$ acting on the modes according to $\hat{q}_j\ket{q}_{\hat{q}_j}=q \ket{q}_{\hat{q}_j}$.  

The ground state is a multimode Gaussian state, which we rewrite for clarity:
\begin{equation}
\ket{G}=\mathcal{N}^{-1}\int dq_0\ldots \int dq_{V-1}\ e^{-\frac{1}{2}\bm{q}^T K^{1/2}\bm{q}}
\ket{q_{0}}_{\hat{q}_0}\ldots \ket{q_{V-1}}_{\hat{q}_{V-1}}
\end{equation}
where $\bm{q}=(q_0,\ldots q_{V-1})^T$ and the coupling matrix $K$ is given in Eq. \ref{Kmatrix}. 
The ground state $\ket{G}$ is obtained by a unitary transformation on the $V$ mode vacuum state, described by the following symplectic transformation on the initially decoupled position and momentum mode operators:
\begin{equation}
\hat{\bm{v}}\rightarrow Y\hat{\bm{v}}.
\end{equation}
The transformation acts to transform the vacuum correlation function as
\begin{equation}
\Gamma_{\rm vac}=\frac{1}{2}\bm{1}_{2V}\rightarrow \Gamma=\frac{1}{2}YY^T,
\end{equation}
where $\Gamma$ is given in Eq. \ref{corrmat}.  Hence the symplectic transformation is:
\begin{equation}
Y=K^{-1/4}\oplus K^{1/4}.
\end{equation}
There is a canonical decomposition for $Y$ written as one round of beam splitters and phase shifters, followed by parallel single mode squeezing, followed by a second round of beam splitters and phase shifters~\cite{Braunstein}. This decomposition is efficient, costing $O(V^2)$ elementary operations.

Particle excitations above the ground state can also be created using the bosonic encoding.  Here the Hamiltonian used to create excitations is a simple quadratic interaction
\begin{equation}
\hat{H}_{\psi}=\hat{f}^{\dagger}\hat{c}+\hat{f}\hat{c}^{\dagger},
\end{equation}
where $\hat{c}^{\dagger},\hat{c}$ are creation and annihilation operators that act on an ancillary bosonic mode.   We prepare the ancillary mode in the Fock state $\ket{n=1}$ and 
evolve by $\hat{H}_{\psi}$, such that $e^{-i\hat{H}_{\psi}\pi/2}\ket{G}\ket{n=1}=-i\hat{f}^{\dagger}\ket{G}\ket{n=0}$ and we have the excited state up to a phase with no entanglement left between the ancilla and the register.  Note the Fock state $\ket{n=1}$ is a non-Gaussian state, however it can be prepared efficiently by a variety of techniques (see~\cite{XBET} and references therein).  
\end{document}